\documentclass{sigchi}

\CopyrightYear{2020}
\setcopyright{acmcopyright}
\doi{https://doi.org/10.1145/3313831.3376367}
\isbn{978-1-4503-6708-0/20/04}
\conferenceinfo{CHI'20,}{April  25--30, 2020, Honolulu, HI, USA}
\acmPrice{\$15.00}

\toappear{\scriptsize 
Permission to make digital or hard copies of all or part of this work for personal or classroom use is granted without fee provided that copies are not made or distributed for profit or commercial advantage and that copies bear this notice and the full citation on the first page. Copyrights for components of this work owned by others than ACM must be honored. Abstracting with credit is permitted. To copy otherwise, or republish, to post on servers or to redistribute to lists, requires prior specific permission and/or a fee. Request permissions from permissions@acm.org. \\
\emph{CHI '20, April 25--30, 2020, Honolulu, HI, USA.} \\
\copyright~2020 Association of Computing Machinery. \\
ACM ISBN 978-1-4503-6708-0/20/04\ ...\$15.00. \\
DOI: \href{http://dx.doi.org/10.1145/3313831.3376367}{http://dx.doi.org/10.1145/3313831.3376367}
}

\usepackage{balance}       %
\usepackage{graphics}      %
\usepackage[T1]{fontenc}   %
\usepackage{txfonts}
\usepackage{mathptmx}
\usepackage[pdflang={en-US},pdftex]{hyperref}
\usepackage{color}
\usepackage{booktabs}
\usepackage{textcomp}

\usepackage{microtype}        %
\usepackage{ccicons}          %

\usepackage{enumitem}
\setlist{nolistsep, topsep=-4pt, partopsep=-4pt}

\usepackage{setspace}
\usepackage{textcomp}
\usepackage{color,soul}
\usepackage{url}
\usepackage[disable]{todonotes}
\usepackage{caption}

\usepackage[final, todonotes={enable}]{changes} 

\usepackage{cuted}
\usepackage{capt-of}

\usepackage{capt-of}

\usepackage{caption}
\usepackage{subcaption}
\def\plaintitle{Tactile Presentation of Network Data: \\ Text, Matrix or Diagram?}

\def\emptyauthor{}
\def\plainkeywords{Accessibility; Blindness; Vision Impairment; Graph Visualization; Adjacency Matrix.}

\makeatletter
\def\url@leostyle{%
  \@ifundefined{selectfont}{
    \def\UrlFont{\sf}
  }{
    \def\UrlFont{\small\bf\ttfamily}
  }}
\makeatother
\def\pprw{8.5in}
\def\pprh{11in}

\definecolor{linkColor}{RGB}{6,125,233}
\hypersetup{%
  pdftitle={\plaintitle},
  pdfauthor={\emptyauthor},
  pdfkeywords={\plainkeywords},
  pdfdisplaydoctitle=true, %
  bookmarksnumbered,
  pdfstartview={FitH},
  colorlinks,
  citecolor=black, 
  filecolor=black,
  linkcolor=black,
  urlcolor=linkColor,
  breaklinks=true,
  hypertexnames=false
}

\begin{document}

\title{\plaintitle}

\numberofauthors{5}
\author{%
  \alignauthor{Yalong Yang\\
    \affaddr{Monash Uni. \& Harvard Uni.}\\
    \affaddr{Melbourne, Australia}\\
    \email{yalongyang@g.harvard.edu}
  }
  \alignauthor{Kim Marriott\\
    \affaddr{Monash University}\\
    \affaddr{Melbourne, Australia}\\
    \email{kim.marriott@monash.edu}
  }
  \alignauthor{Matthew Butler\\
    \affaddr{Monash University}\\
    \affaddr{Melbourne, Australia}\\
    \email{matthew.butler@monash.edu}
  }
  \alignauthor{Cagatay Goncu\\
    \affaddr{Monash University}\\
    \affaddr{Melbourne, Australia}\\
    \email{cagatay.goncu@monash.edu}
  }
  \alignauthor{Leona Holloway\\
    \affaddr{Monash University}\\
    \affaddr{Melbourne, Australia}\\
    \email{leona.holloway@monash.edu}
  }
}

\maketitle

\begin{abstract}
  Visualisations are commonly used to understand social, biological  and  other kinds of networks. Currently we do not know how to effectively present network data to people who are blind or have low-vision (BLV). We ran a controlled study with 8 BLV participants  comparing four tactile representations: organic node-link diagram, grid node-link diagram, adjacency matrix and braille list. We found that the node-link representations were preferred and more effective for path following and cluster identification while the  matrix and list were better for adjacency tasks. This is broadly in line with  findings for the corresponding visual representations. 
\end{abstract}




\section{Introduction}
Networks are ubiquitous~\cite{lima2007visual}. Visualisations such as node-link diagrams or matrices are widely used to understand networks. They underpin how the theory of graphs and networks is taught at school and university, they are commonly used in the analysis of social networks,  biological networks and software, and they are used to represent networks in popular media\todo{add references}.  However, if you are blind or have low-vision (BLV) such visual representations are not accessible. 

Current accessibility guidelines recommend the use of raised line drawings called tactile graphics to present graphics for which spatial relationships are important, such as maps or mathematical graphs~\cite{BANA2010}.  Thus a natural question to ask is: What is the best way to represent a network in a tactile graphic? This is the question we address here. As far as we can tell we are the first to do so. While there has been research into how to represent standard statistical graphics such as bar charts, pie charts and scatter plots~\cite{goncu2010usability,watanabe2018effectiveness}, there has been no research into the tactile representation of networks. Given the widespread use of networks, it is of significant importance to investigate this for BLV users.

\begin{figure*}[t!]
  \centering
  \includegraphics[width=0.85\textwidth]{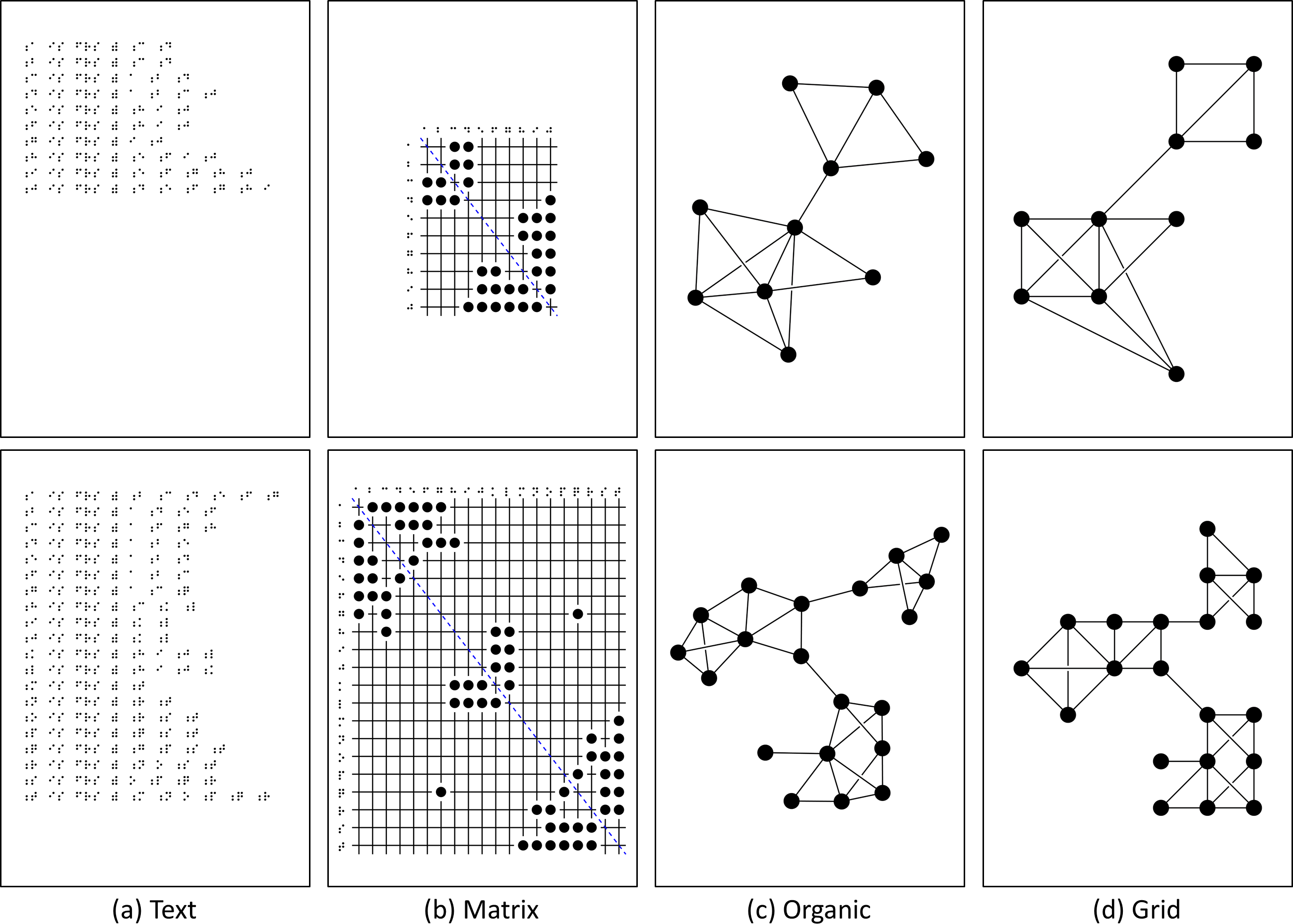}  
  \caption{Two of the tested stimuli in different tactile representations: top with 10 nodes and bottom with 20 nodes.}
  \label{fig:teaser}
\end{figure*}

We conducted a controlled user study with eight blind participants who were all touch readers. We compared four representations. These are shown in Figure~\ref{fig:teaser}.

The first was a node-link diagram laid out in an organic style (see~\autoref{fig:teaser}(c)). Such layouts are produced by force-directed graph drawing algorithms and are the most common way of visualising a network~\cite{battista1998graph}. %
Touch readers must explore a graphic element by element, slowly building up a mental model of the graphic. 
In organic layouts, the placement of nodes and orientation of edges appears quite haphazard. We felt that this might make it difficult for touch readers  to build a mental model and that they might confuse nodes and not recognise that they had already visited a node when arriving by a different path. 

Our second representation was also a node-link diagram, but this time laid out on a grid (see~\autoref{fig:teaser}(d)). This means that the location of nodes is much more regular and edge orientations are more consistent, typically vertical or horizontal. There is some evidence that sighted readers prefer grid-layouts~\cite{purchase2010graph} and that they are more memorable than organic layouts~\cite{marriott2012memorability}. We conjectured that touch readers would find these diagrams easier to navigate and that they would find it easier to build an accurate mental model of the network.

Our third representation was a matrix (see~\autoref{fig:teaser}(b)). A number of studies have found that for sighted readers,  matrix representations are more effective than node-link diagrams for most tasks using larger, more cluttered graphs~\cite{ghoniem2004comparison,okoe2018node}. We conjectured that the benefits of matrix representations might be even greater for touch readers. The organisation is predictable and we thought the tabular structure would be easy to navigate by touch. Furthermore they do not contain edge crossings, which we thought might confuse touch readers. The one task for which node-link diagrams offer advantages over matrix representations is for finding a path between two nodes. Here the position of the target node relative to the source node helps to guide the choice of which edges to follow~\cite{huang2007using}. We were curious to see if this advantage also holds for touch readers.

Our final representation was simply a list giving the neighbours for each node (see~\autoref{fig:teaser}(a)). This representation was designed to be a baseline to test whether there are in fact benefits to using a graphical representation for networks over a purely textual representation. Text descriptions are commonly used as representations of graphical content for BLV people. 

As we expected we found the two node-link representations generally out performed the other two representations and were found to be more intuitive. However we did not find performance or a preference for the grid layout over the organic style. We were also surprised by the poor performance of the matrix representation, which in most cases was similar to or worse than that of the text representation.

This research contributes to a more nuanced understanding of the relative advantages of text and graphics to BLV people. It also provides accessible format providers with guidelines on how to appropriately transcribe network data in educational and other materials. 
\vspace{-1em}
\section{Related Work}
We review previous findings on visual presentation of network data and tactile graphic design.

\vspace{-0.5em}
\subsection{Visualisation of Network Data}
Networks have been visualised since the Middle Ages. However, it is in the mid-twentieth century that networks began to be used in the social sciences, life sciences, information technology and engineering to model complex relational data~\cite{lima2007visual}. Shortly afterwards, computer scientists began to develop automatic graph layout algorithms for visualising networks~\cite{battista1998graph}. 

Most networks are visualised using some form of node-link diagram. One of the first methods to network layout was force-directed layout~\cite{eades1984heuristic}. These treat the graph as a physical system in which nodes repel one another but edges are like springs, pulling connected nodes together. Force-directed layouts give rise to node-link diagrams laid out in  an organic-style  in which edges have similar lengths, clusters are separated and to a great extent the graph is untangled to reduce edge crossings. Since then many algorithms have been proposed for drawing  organic-style layouts~\cite{gibson2013survey} and such layouts remain the most popular way of visualising networks with undirected edges.

Other layout styles for node-link diagrams have also been investigated. Popular methods are layered diagrams for directed graphs, chord diagrams and orthogonal layouts in which the edges are composed of horizontal or vertical line segments~\cite{battista1998graph}. More recently, algorithms for creating grid-like layouts in which nodes are placed on a grid and connected by horizontal, vertical or diagonal edges have been developed~\cite{kieffer2015hola}. 

However, regardless of the layout style, node-link diagrams become cluttered and difficult to understand when the graph is dense and there are many edge crossings. For this reason visualisation researchers have investigated adjacency matrix-based representations, e.g.~\cite{abello2004matrix,henry2007nodetrix,ghoniem2004comparison}. The rows and columns are labelled by the nodes and if there is an edge between two nodes then the cell  whose  row and column is associated with the nodes is shaded. Automatic layout algorithms reorder the rows and columns so that the rows/columns of connected nodes are adjacent in the matrix~\cite{Fekete2015ReorderjsAJ}. Clusters are revealed as rectangular blocks along the diagonal. Studies comparing node-link and matrix representations have found that for larger graphs many tasks (estimating graph size, finding most connected node, finding common neighbour) are faster and/or more accurate with a matrix representation ~\cite{ghoniem2004comparison,keller2006matrices,okoe2018node}. However, path finding tasks are more difficult with matrix representations.

\subsection{Tactile Graphics}
The most common production methods for tactile graphics are embossing with raised dots using a braille embosser, printing onto swell paper and thermoforming~\cite{rowell2003world}.
Swell paper contains microcapsules of alcohol. When the paper is heated, areas printed with carbon  swell to an elevation of around 0.5mm. 
Preferences vary but a survey of 30 adults with severe vision impairment showed a preference for maps printed on swell paper over thermoform or (handmade) models~\cite{rowell2005feeling}.

Accessibility guidelines give general advice on choice of format for access to print graphics via tactile graphics, audio description or models~\cite{ABS2002, DIAGRAM2018, Gale2005}. For tactile graphics, recommendations are given on design aspects such as simplification, spacing and fill patterns~\cite{BANA2010}. Specific guidelines are given for some common charts, but we could not find these for network diagrams. In practice, transcribers follow the print format except in rare cases when an alternative is suggested by a teacher or author. Evidence is required before transcribers will incorporate presentation mode changes into the transcription process. The current study helps to provide such evidence.

Some studies have compared the effectiveness of tactile graphic and tabular representations of data. One study compared a tactile table with a tactile bar chart for a small data set with 10 points~\cite{goncu2010usability}. Participants preferred the tactile table. Another study compared a tactile table, a digital text file and tactile scatter plots with 20 data points in respect to identifying the data trend quickly and correctly~\cite{watanabe2012development,watanabe2018effectiveness}. Here, the tactile scatter plot outperformed the other representations. An early study~\cite{aldrich1987tangible} compared different formats for tactile line charts. Another study~\cite{engel2018user} evaluated different tactile formats for bar charts, line charts, pie charts and scatter plots.
To the best of our knowledge, there has been no research evaluating the effectiveness of different tactile representations of networks.

It is worth pointing out that differences between haptic and visual perception mean that the advantages found for a particular graphic formalism presented visually may not hold when presented tactually. The main difference is the limited perceptual area  of touch when compared to vision~\cite{schwartz2017sensation}. Touch readers must sequentially explore the elements in the graphic to build up a mental model of the spatial layout. They are taught strategies for systematically exploring a tactile graphic in order to build an overview~\cite{Berla1973,picard2012}. 

While prior research has not explored tactile representations for network diagrams, they have explored accessible haptic and audio representations. For instance, TeDub allowed the user to explore a UML diagram with a phantom haptic device guiding the exploration~\cite{Horstmann2004}. 
Other projects used keyboard navigation and audio labels to enable self-creation of node-link diagrams and collaboration between BLV and sighted users~\cite{Balik2013, Balik2014, Matti2017}, however the focus was autonomy rather than effectiveness for understanding.

\section{User Study}
Our main contribution is a controlled study comparing different tactile representations of network data with BLV people. We couched the study as showing social network data because social network visualisations are common in educational materials and popular media and because the underlying concepts of network data analysis are readily explained in this context to people who may be unfamiliar with networks. 
We considered four representations:
\begin{itemize}
	\item Text (Txt for short): Braille list detailing the friends of each person. This provided a purely textual representation as a baseline for comparison with the three diagrammatic representations. 

	\item Matrix (Mat for short): Adjacency matrix-based representation. This has been found to be more effective than a node-link representation for many tasks for sight readers. We hypothesized that it would provide similar benefits for touch readers. 

    \item Organic (Org for short): Node-link diagram laid out in an organic style. This is the most common visual representation of network data.

    \item Grid (Grd for short): Node-link diagram laid out in a grid style. We hypothesized that placing the nodes on grid points would improve the readability of the node-link diagram for touch readers.
\end{itemize}

\subsection{Medium}
We tested both swell paper and thermoform tactile representations with an experienced touch reader. 
They reported difficulties in interpreting the links associated with a high degree node (\textit{i.e.} a node with many links), on the thermoform representation and in general preferred swell paper. This preference is in accord with~\cite{rowell2005feeling}. We therefore decided to use swell paper  in the user study. We used A4 size.

\subsection{Network Data}
Two sizes of network data were tested in the user study: \emph{small} with 10 nodes and \emph{large} with 20 nodes. While still limited in size, these are representative of the kinds of networks used in educational materials. We also varied the number of clusters in each data set: For the small size, a network could have either 1 or 2 clusters; for the large size, a network could have either 2 or 3 clusters. A stochastic block model~\cite{holland1983} was used to generate network data. We created 8 data sets for trials and 1 for training. 36 (9 data sets$\times$4 representations) tactile graphics on swell paper were produced.

\subsection{Representations}
The representations were developed in consultation with an experienced transcriber of tactile materials and tested with an experienced  touch reader. Examples are shown in \autoref{fig:teaser}.

\begin{figure}
	\centering 
	\includegraphics[width=0.8\columnwidth]{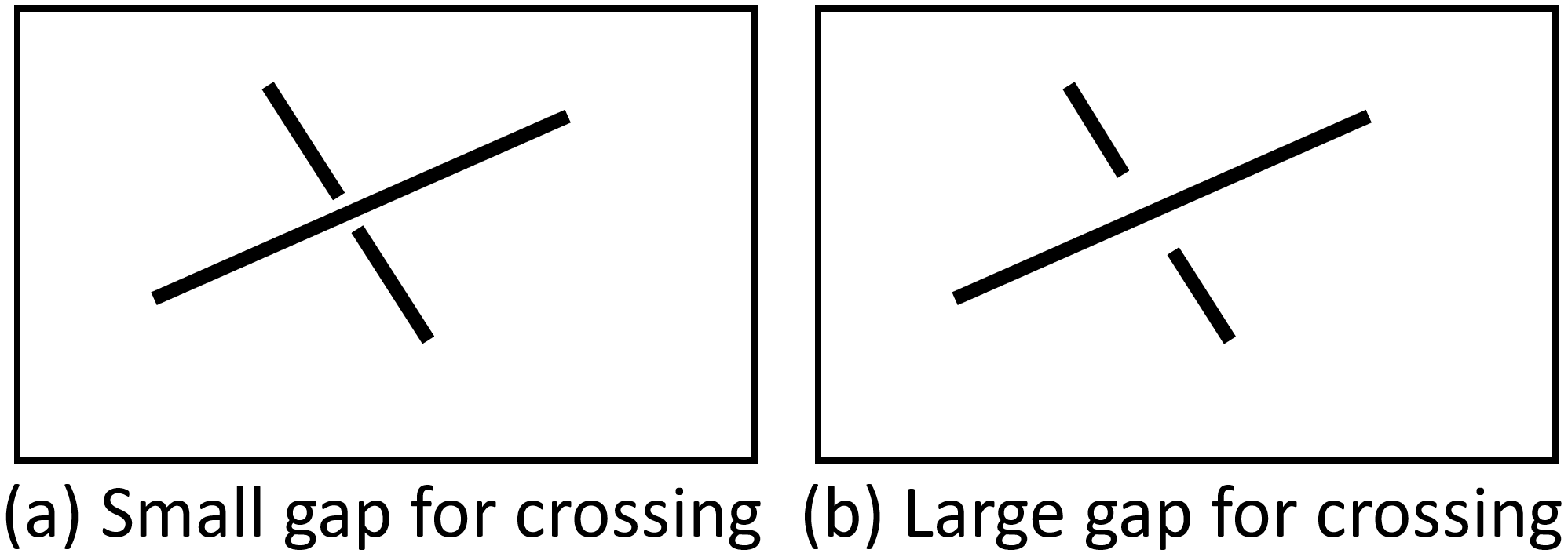}
	\caption{Gaps for link-link crossing on swell paper.}
	\label{fig:crossing-gap}
\end{figure}

\textbf{Node-link diagram:} 
\textbf{\emph{1) Layout:}} We used yEd~\cite{yed}, a widely-used industry standard graph layout tool, to create organic node-link diagrams. Based on the organic node-link diagrams, we manually manipulated the positions of the nodes to create the grid node-link diagrams. We kept the overall layout and edge length in the grid style node-link diagrams similar to the original organic node-link diagrams so as to reduce confounds. 
\textbf{\emph{2) Crossings:}}  Edge crossings are common in node-link diagrams. The experienced transcriber recommended that we introduce gaps to allow touch readers to distinguish the individual edges, like \autoref{fig:crossing-gap}. We tested both small and large gaps with an experienced touch reader, who preferred the small gap as \textit{``it is easier to follow''}. Therefore we used a small gap in the user study.
\textbf{\emph{3) Size:}} The nodes were 30~pt in diameter and the line thickness was 2.25~pt.

\begin{figure}
	\centering 
	\includegraphics[width=0.8\columnwidth]{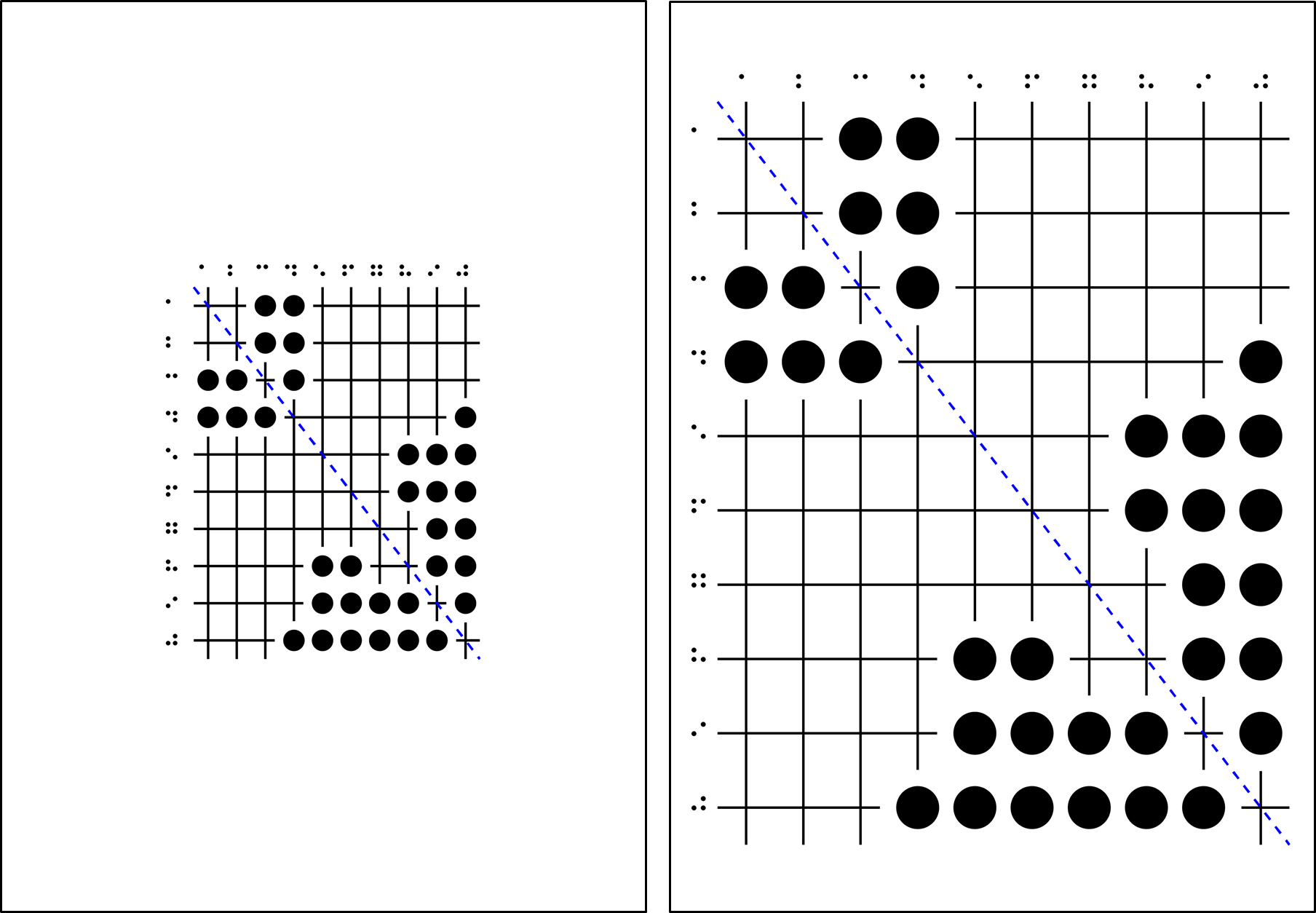}
	\caption{Matrix for small network (10 nodes) in different sizes on A4 swell paper.}
	\label{fig:matrix-size}
\end{figure}

\textbf{Matrix:} 
\textbf{\emph{1) Order:}} When visualizing a network as a matrix, a good order of rows/columns can reveal high-level patterns in the data~\cite{Fekete2015ReorderjsAJ}. To produce a spatially coherent and well organized matrix, we followed the recommendation in~\cite{behrisch2016} and used the \emph{Optimal-Leaf-Ordering} algorithm to calculate the order of rows/columns. We kept matrix symmetry, \textit{i.e.} the rows and columns have the same order.
\textbf{\emph{2) Braille labels:}} Braille labels were not needed for the nodes in the node-link representations and our experienced transcriber recommended omitting them to reduce clutter. By contrast, braille labels are required for the matrix representation because the reader needs to know which rows and columns correspond. We used a single braille letter on the left of the matrix to label the rows and at the top of the matrix for columns. The braille letters were in alphabetic order. We used standard braille with a 24~pt font size.
\textbf{\emph{3) Aspect ratio:}} We laid the matrix out in portrait mode as braille standards allow the horizontal spacing between letters to be smaller than vertical spacing between the rows and if the large size graph (20 nodes) was laid out in landscape mode the spacing between the rows was insufficient. We tested aspect ratio of 1, 1.1, 1.2 and 1.3 with the experienced touch reader. The braille rows in the matrix with aspect ratio of 1.2 and 1.3 were readable, however the touch reader preferred the extra space of the 1.3 aspect ratio. 
\textbf{\emph{4) Grid lines:}} Based on the advice of the experienced transcriber instead of having grid lines separating each row/column, we used one line through the center of each row/column to better guide a finger along the row/column. The thickness of the line was 2.25~pt.
\textbf{\emph{5) Circle size:}} A circle was placed at the crossing of a row and a column when the two associated nodes are connected in the network. The large size network allowed a maximum diameter of 20~pt. We used the same size for the small network.
\textbf{\emph{6) Matrix size:}} For the small network, we were also able to make the matrix fill the full paper (see~\autoref{fig:matrix-size}). We tested both options in \autoref{fig:matrix-size} with the experienced touch reader. The touch reader did not like the circles to be as large as those in right sub-figure of \autoref{fig:matrix-size} and reported it to be more difficult to retrieve information. Thus we used the more compact layout.
\textbf{\emph{7) Diagonal line:}} We used a spur wheel to mark the diagonal line with a texture distinct from the other lines.

\textbf{Text:} 
We used the same order of rows/columns in the matrix for the braille list. There was one row for each letter in the form ``X is friends with U V ...''
We used a standard 24~pt font size for the braille.

\subsection{Experiment}
\textbf{Tasks:}
Based on previous user studies comparing node-link and matrix visualizations~\cite{ghoniem2004comparison,okoe2018node} and task taxonomy of network visualisation~\cite{Lee:2006cs}, we identified six essential tasks when exploring network data. We focused on graph topology and overview tasks as these are the most fundamental tasks and the ones in which we might expect graphical representations to offer benefits over a braille list.  In order for target nodes in the tasks to be found quickly by touch readers we ``highlighted'' the nodes by placing a piece of sticky plasticine on them and gave verbal directions on where to find them on the presentation, e.g. ``in the top left corner.''  
\begin{itemize}
	\item T1: Estimate the number of friendship links.\footnote{Overview tasks usually include estimating the number of nodes. Because of the Braille labelling of the list and matrix it did not make sense to estimate the number of nodes since Braille letters are also Braille for numbers, so the rows were actually numbered in Braille from 1 to 10 or 1 to 20 depending upon the size} (Overview)
	\item T2: How many friendship groups are in this social network? (Topology-connectivity (cluster))
	\item T3: Find the two people with sticky plasticine. Are they friends? This was repeated. (Topology-adjacency)
	\item T4: Find the three people with sticky plasticine. How many friends do they each have? (Topology-adjacency)
	\item \added{T5: Find the two people with sticky plasticine. Who are their mutual friends? (Topology-common connection)}
	\item \added{T6: Find the two people with sticky plasticine. Now find a path of direct friends connecting them and show us the path. (Topology-connectivity path)}
	
\end{itemize}

\textbf{Set-up:}  Participants were seated comfortably in front of a desk and the tactile graphics were placed on top of the desk. A video camera was placed to capture hand movements of participants and to record their answers. They were tested in a location of their choice.

\textbf{Participants:} We recruited eight participants (five female and three male) through email lists and personal contacts. Six of the participants identified themselves as totally blind and two as legally blind.
All were experienced touch readers and familiar with braille. Six participants rated themselves \emph{expert} in using tactile diagrams and the other two rated themselves \emph{knowledgeable}. Three (P2, P3, P8) had previously used networks 
but not seen the graphical representations. %

\textbf{Design and Procedure:} We used a within-participant design
with each participant performing tasks with each representation. The order of presentation was balanced using a Latin square design. \added{For each participant, the networks were different in the different conditions, except for the training material in which the same network data was used.}

Participants were first given a brief introduction to the experiment. They were then presented with the four representations. The process for each representation was as follows. 

First the participant was trained in its use. They were asked to explore it and were explained concepts.
\added{We deliberately did not use technical terminology in the training. For each representation, we used friendships to explain the concepts. We first explained what represents a person in the presentation, \textit{e.g.} in a matrix, each row or column stands for a person. We then explained how a friendship between two people was represented, \textit{e.g.} in a matrix, if there is a circle at the intersection of person A, a row or a column, and person B, a column or a row, then it means A and B are friends. We emphasised that in a matrix a friendship is represented by two symmetrically placed circles, and that in the text representation, the friendship is recorded twice. Finally, we explained a friendship group is a group of people who are friends with people within the group but not with people outside the group and that only one person in a friendship group can be a friend with a person outside of the group.}

We then asked them to complete sample tasks from T2 to T6. We did not teach explicit strategies but endeavoured to ensure they understood the goal of the tasks. We did not train in T1 as we were interested in their estimation after obtaining a general overview rather than after explicitly trying to determine the number of links. 
Participants were encouraged to ask questions in the training and corrected if wrong.

They were then asked to complete the tasks with the representation.
Originally we planned to ask participants to complete all tasks with the two data sets (10 and 20 nodes). Pilot testing showed this took too long. Instead, we first showed them the small data set and asked participants to explore it. We then asked them to complete tasks T1 and T2. We then showed them the larger data set and asked them to explore it and then asked the participants to complete all tasks (T1 to T6). This allowed us to obtain two measurement for T1 and T2 and meant that all tasks were completed on the network that we felt was of a more representative size.

Time for exploration and the time for each task and answer were recorded. After each task, participants were asked to rate its difficulty.  Following~\cite{huang2009measuring,bratfisch1972perceived}, we used a nine-grade symmetrical category scale: very very easy; very easy; easy; rather easy; neither easy nor difficult; rather difficult; difficult; very difficult; and very very difficult. 
Participants were told explicitly that they could give up if they struggled to complete a task and that the difficulty would be recorded as ``very very difficult''. We also asked the strategies they used to answer the various tasks. 

After completing the tasks for a representation participants were asked to indicate their agreement on a 5-point scale from strongly disagree to strongly agree, with each of the following statements:
\begin{itemize}
	\item S1: The representation was easy to understand and I barely needed instruction in how to use it.
	\item S2: I imagined the social network in my head.
\end{itemize}
If they could imagine it they were asked to elaborate on what it looked like.

After completing the tasks for all four representations, we asked the participant to rank the representations. We presented all representations in the training data set and asked them to put them in an order such that the preferred one was on the left. We asked the ranking for:
\begin{itemize}
	\item Q1: The most natural or intuitive way of showing a social network
	\item Q2: In general, the easiest to use
\end{itemize}
They were then invited to provide comments explaining their ranking.
Finally we collected demographic data and asked if they had previously seen any of the representations. The study lasted one hour and forty minutes on average.

\section{User Study Results and Discussion}
\subsection{Data analysis}
Times were log-normal distributed (checked with histograms and Q---Q plots) in all tasks. We used \emph{linear mixed modeling} to evaluate the effect of independent variables on times~\cite{Bates2015}. All independent variables and their interactions were modeled as fixed effects. A within-subject design with random intercepts were used for all models. The significance of inclusion of an independent variable or interaction terms were evaluated using log-likelihood ratio. Tukey's HSD post-hoc tests were then performed for pair-wise comparisons using the least square mean~\cite{Lenth2016}. Homoskedasticity and normality of the Pearson residuals were evaluated graphically using predicted vs residual and Q---Q plots respectively. Degree of freedom, $\chi^2$ and $p$ value for fixed effects were reported following~\cite[p.~601]{field2012discovering}. Time and log transformed time with 95\% confidence of different representations are reported using figures.

For error rate, participants completed T3, T4, T5 and T6 almost all correctly. We report case by case in different tasks when participants gave us a wrong answer or they gave up trying. For T1 and T2, we used Friedman test to compare the error rate among different tactile representations.

For self-reported difficulty (short for difficulty), feedback rating and rankings, we again used Friedman test to compare different tactile representations

We also reported the detailed collected data in figures along with aggregated figures.

\subsection{T1: Overview (estimate number of links)}
All participants used both hands to systematically explore all representations. In the matrix and text representation, participants often followed the rows, while in the node-link diagrams they  followed the links.
With the text representation P4 and P7 indicated information overload: 
\begin{quote}
   \emph{``too many people, can't get into my head'' [P4]} 
\end{quote}
 
\begin{table}[t!]
    \centering 
    \includegraphics[width=\columnwidth]{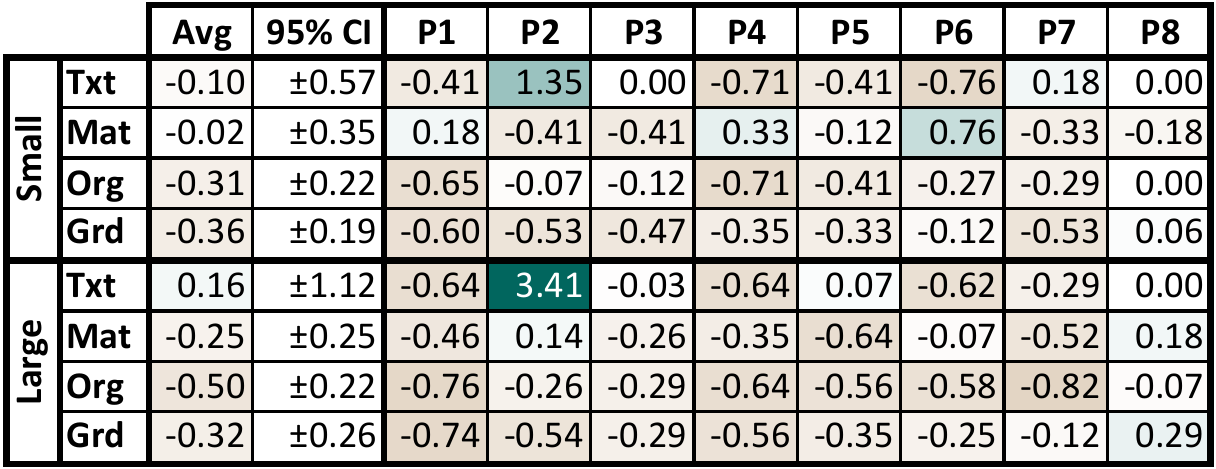}
    \caption{Error rate for T1: aggregated 95\% confidence interval and error rate per participant.}
    \label{tab:error-t1}
    \vspace{0.3em}
\end{table}
\begin{table}[t!]
    \centering 
    \includegraphics[width=\columnwidth]{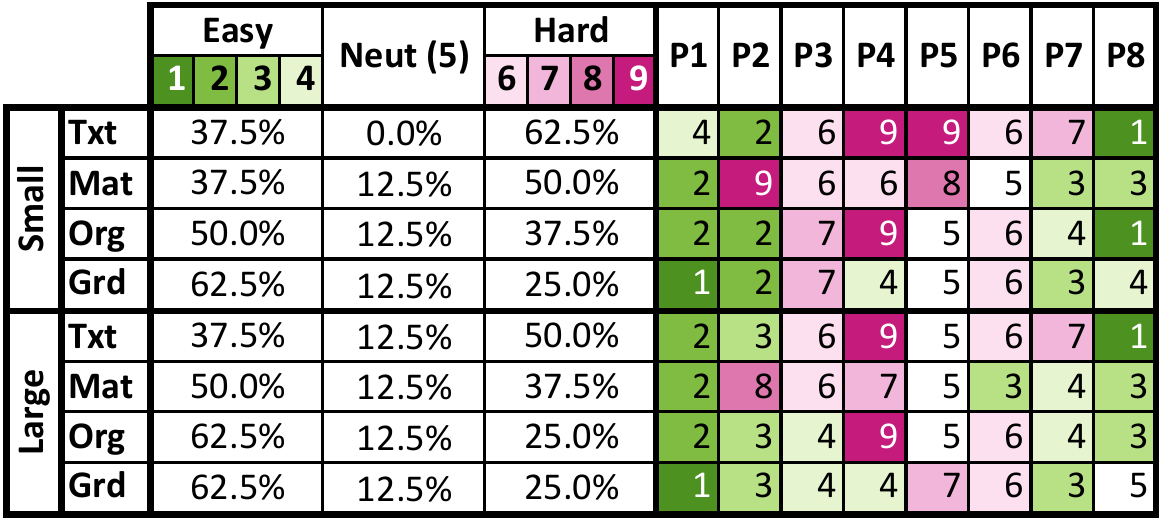}
    \caption{Difficulty for T1: aggregated percentage of difficulty ratings and difficulty ratings per participant.}
    \label{tab:difficulty-t1}
\end{table} 
 
We used~\autoref{equ:T1-error} to measure the difference between the estimated number of links and the correct answer. 
 \begin{equation}
errorRate = \frac{ParticipantAnswer - correctAnswer}{correctAnswer}
\label{equ:T1-error}
\end{equation}
The error rate was high but we did not find any statistically significant difference between the representations (see \autoref{tab:error-t1}).
In the node-link representations participants consistently underestimated the number.  The reason may be that the lines representing links are not tactually salient: when first exploring the representation many are simply not noticed. 

The difficulty ranking for the task varied greatly between participants (see \autoref{tab:difficulty-t1}). However, at least one participant (P1) said that it was easy because she was simply guessing. P4 and P6 also said that they guessed.  Participant P2 used the number of rows (which he knew from the braille labeling) in the matrix and text representation and multiplied this by his estimate of the average node degree to compute an estimate of the number of links. 

Overall, estimating the number of links in all representations seems considerably more difficult for touch readers than for sighted readers with the equivalent visual representations. There was no obvious performance difference between participants with experience of using network data and participants without such experience.

\begin{table}[t!]
    \centering 
    \includegraphics[width=0.73\columnwidth]{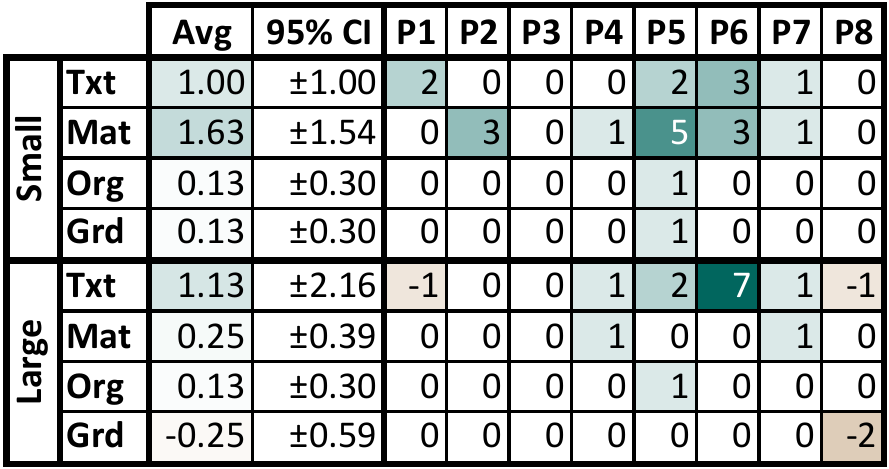}
    \caption{Error rate for T2: aggregated 95\% confidence interval and error rate per participant.}
    \label{tab:error-t2}
    \vspace{0.3em}
\end{table}
\begin{table}[t!]
    \centering 
    \includegraphics[width=\columnwidth]{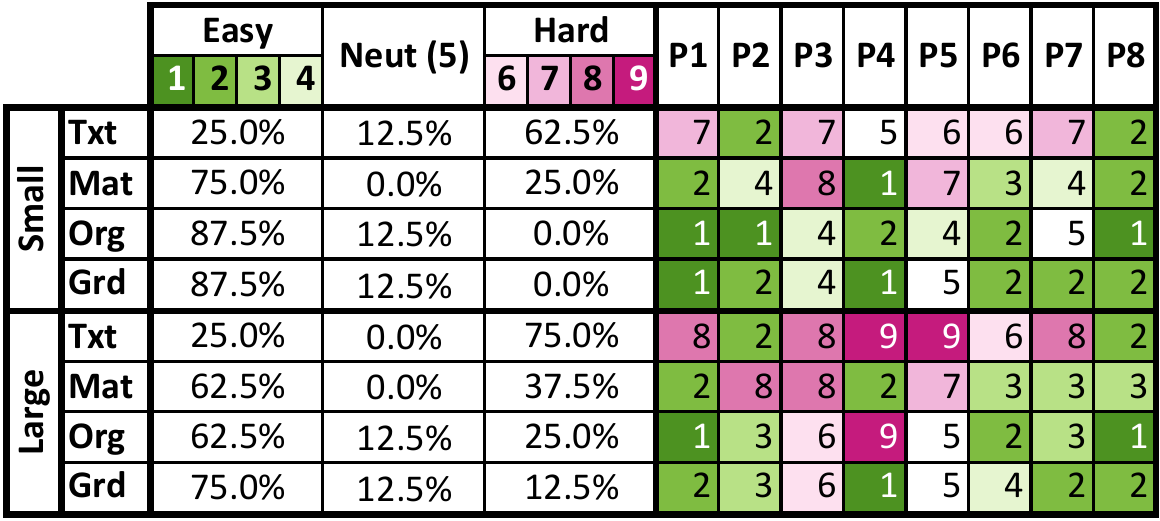}
    \caption{Difficulty for T2: aggregated percentage of difficulty ratings and difficulty ratings per participant.}
    \label{tab:difficulty-t2}
\end{table}

\subsection{T2: Connectivity-cluster (number of clusters)}
When analysing the error rate for determining the number of clusters we used $ParticipantAnswer - correctAnswer$.

Participants clearly found it more difficult to determine the number of clusters with the text and matrix representations than with the node-link diagram representations (see \autoref{tab:error-t2}). \added{Friedman test revealed a statistically significant effect of representations on error rate in small data ($\chi^2(3)=11.9, p=.0076$) and large data ($\chi^2(3)=8.9, p=.0302$). Post-hoc tests revealed that:}
\begin{itemize}
    \item Grid was more accurate than Matrix (Small: $p=0.0285$; Large: $p=0.0723$) and 
    \item Organic was more accurate than Matrix (Small: $p=0.0285$; Large: $p=0.0368$).
\end{itemize}

\begin{table*}
    \centering 
    \includegraphics[width=\textwidth]{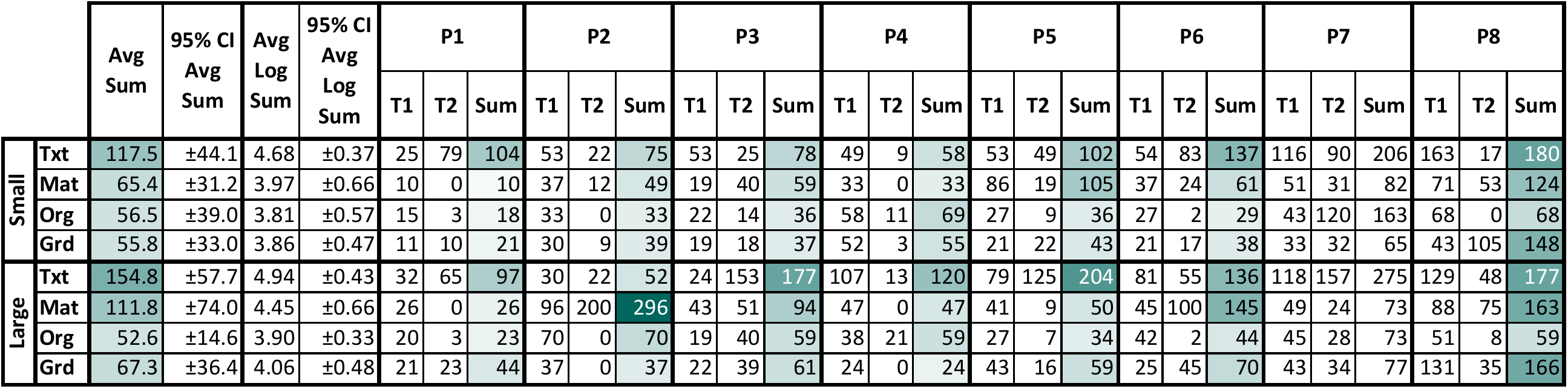}
    \caption{Time for T1, T2: aggregated 95\% confidence interval and time per participant.}
    \label{tab:time-t1-t2}
\end{table*}

This is line with the self-reported difficulty, which was also higher with the text and matrix though only statistically significant for text (see \autoref{tab:difficulty-t2}). \added{Friedman test only revealed a statistically significant effect of representations on difficulty for the small data set($\chi^2(3)=12.6, p=.0056$). Post-hoc tests revealed that:}
\begin{itemize}
    \item Grid was easier than Matrix (Small: $p=0.0380$) and 
    \item Organic was easier than Matrix (Small: $p=0.0209$).
\end{itemize}

This is also supported by an analysis of participant time. Because participants were aware that they were going to be asked about the number of clusters in the network several indicated that they also considered this question when initially exploring the network. Therefore a better indicator of the actual time taken to determine the clusters is the time taken for this task plus the time for initial exploration.  We see that participants took significantly longer exploring the network and determining the number of clusters with the text  than with the other three representations (see \autoref{tab:time-t1-t2}). \added{Linear mixed modeling revealed a statistically significant effect of representations on the total time ($\chi^2(3)=30.6, p<.0001$).} Post-hoc tests revealed that:
\begin{itemize}
    \item Grid was faster than Text ($p<0.0001$),
    \item Organic was faster than Text ($p<0.0001$),
    \item Matrix was faster than Text ($p=0.0044$) and
    \item In linear mixed modeling, we modeled the data size as a factor and found participants were faster with small data than large data ($p=0.0293$)
\end{itemize}

Analysis of the strategies revealed that all participants except P4 used spatial strategies for determining the number of clusters in the node-link representations: using two hands to find possible clusters of nodes and links and then identifying if there was a bridge between the clusters.

Participants generally used a spatial strategy for finding  clusters in the matrix representation: looking for rectangular groups. One participant mentioned also looking for bridges (P3). Some (P2, P7) were  confused by the symmetry, incorrectly identifying symmetric blocks of dots as different clusters.

The strategies for identifying clusters in the textual representation were less defined and some participants indicated that they read the braille and then made an educated guess (P3, P4) while P1 said she could not work out how to find clusters. Others tried to build a representation in their head: P2 explicitly tried to build a matrix representation as he had just seen that in the previous trials. Others indicated the difficulty of keeping it in their head: \begin{quote}
  \emph{``It's overwhelming for me to think about all that.'' [P6]}  
\end{quote}

There was no obvious performance difference between participants with experience of using network data and participants without such experience.

\vspace{-0.5em}
\subsection{T3: Adjacency (are they connected)}
All participants could determine if two nodes were connected by a link correctly for all representations. However, participants were significantly slower with the matrix representation and slightly slower with the text representation than with the two node-link diagram representations (see \autoref{tab:time-t3}). \added{Linear mixed modeling revealed a statistically significant effect of representations on time ($\chi^2(3)=48.3, p<.0001$). Post-hoc tests revealed that:} 
\begin{itemize}
    \item Grid was faster than Text ($p=0.0004$) and Matrix($p<0.0001$),
    \item Organic was faster than Text ($p=0.0002$) and Matrix($p<0.0001$), and
    \item Text was faster than Matrix ($p=0.0727$).
\end{itemize}

The performance on time was in accord with self-reported difficulty though the difference was only statistically significant for the matrix representation (see~\autoref{tab:difficulty-t3}). \added{Friedman test revealed a statistically significant effect of representations on difficulty for T3 ($\chi^2(3)=9.1, p=.0286$). Post-hoc tests revealed that:}
\begin{itemize}
    \item Grid was easier than Matrix($p=0.0901$) and
    \item Organic was easier than Matrix($p=0.0901$)).
\end{itemize}

\begin{table}[t!]
    \centering
    \includegraphics[width=\columnwidth]{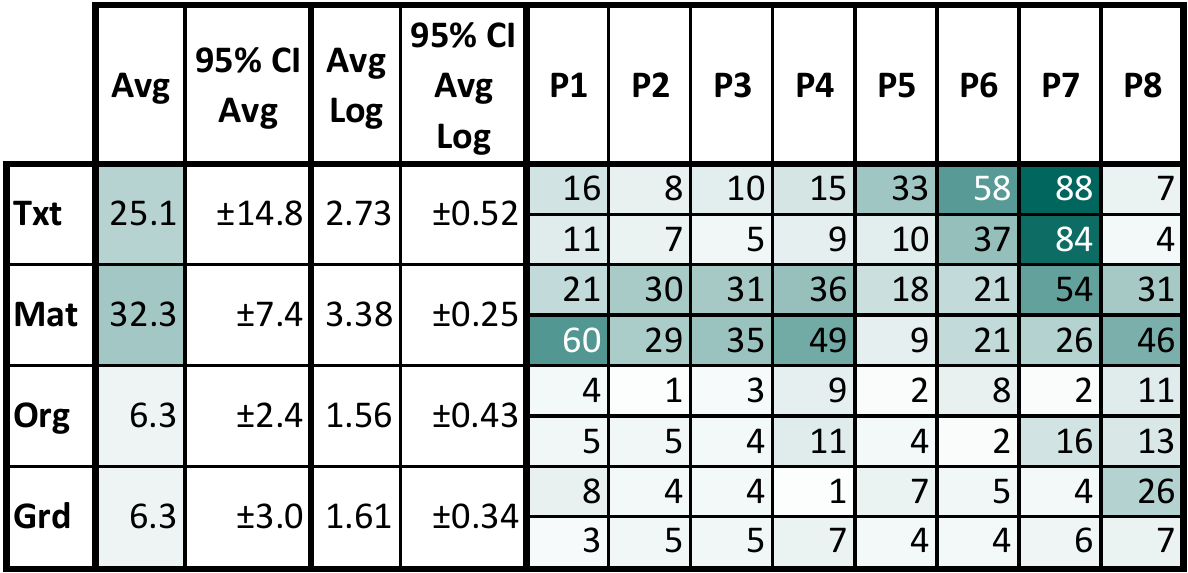}
    \caption{Time for T3: aggregated 95\% confidence interval and time per participant.}
    \label{tab:time-t3}
\end{table}
\begin{table}[t!] 
    \centering 
    \vspace{0.2em}
    \includegraphics[width=0.9\columnwidth]{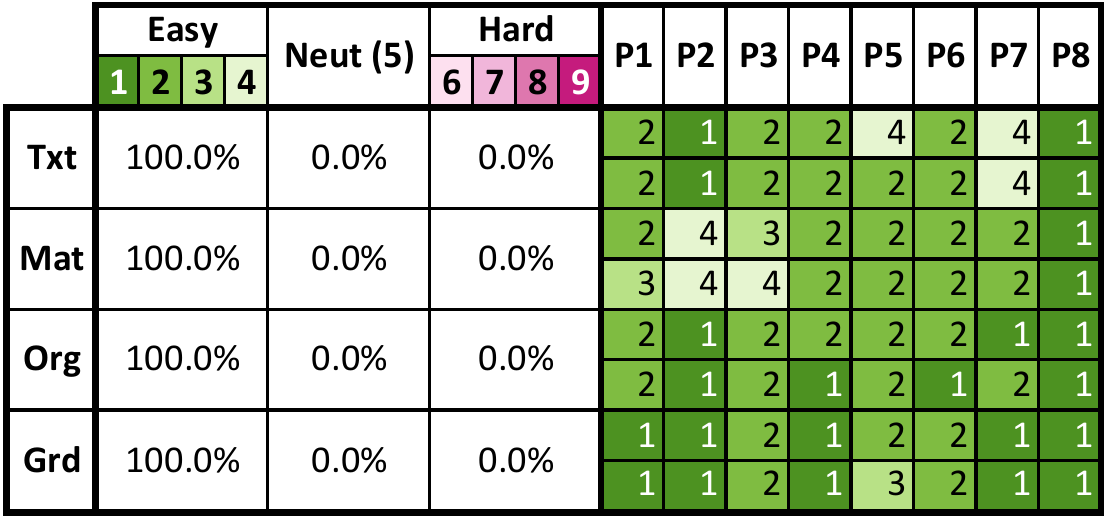}
    \caption{Difficulty for T3: aggregated percentage of difficulty ratings and difficulty ratings per participant.}
    \label{tab:difficulty-t3}
\end{table}

Strategies were similar to those used by sighted readers when using the visual equivalent. In the text representation participants would read along the row to see if the other name was mentioned. For the matrix representation most used two hands, one tracing across the row, one tracing down the column to find if there was a circle in the intersecting cell. Participants P1 and P2 mentioned the matrix was a bit cluttered making it difficult to follow the row and column. One participant suggested adding labels to the right-hand side and bottom of the matrix. 
For the node-link representation participants checked if there was a line between the two nodes.  
One participant initially didn't realise that a node ``breaks'' the link and so was confused by contiguous horizontal edges in the grid layout.

Participants with experience of using network data ($M=6.8s, SD=2.1$) were faster and performed more consistently with Text than the participants without experience ($M=36.1s, SD=30.5$).

\begin{table}[t!]
    \centering 
    \includegraphics[width=\columnwidth]{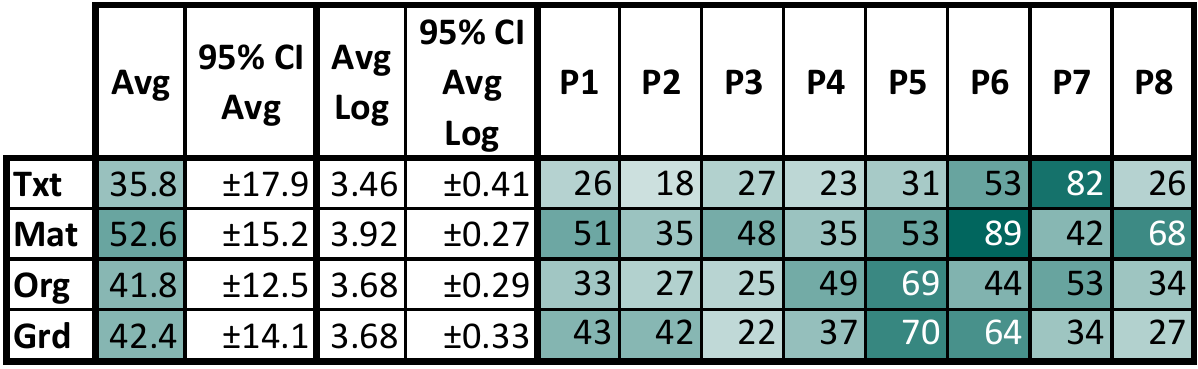}
    \caption{Time for T4: aggregated 95\% confidence interval and time per participant.}
    \label{tab:time-t4}
    \vspace{0.4em}
\end{table}
\begin{table}[t!]
    \centering 
    \includegraphics[width=0.9\columnwidth]{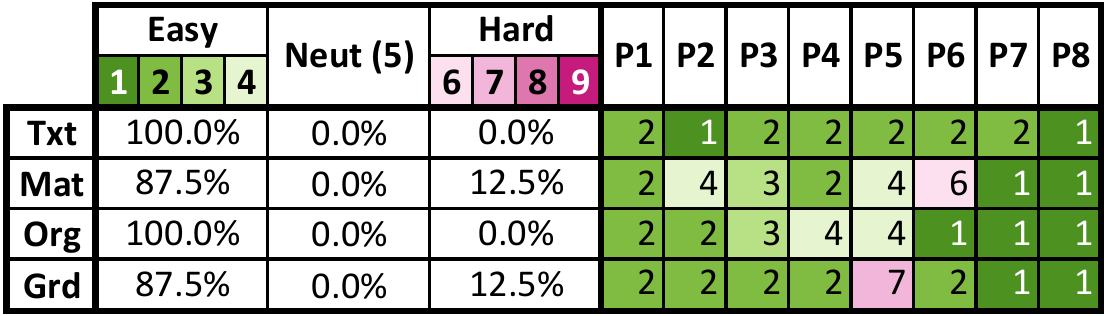}
    \caption{Difficulty for T4: aggregated percentage of difficulty ratings and difficulty ratings per participant.}
    \label{tab:difficulty-t4}
\end{table}

\vspace{-0.5em}
\subsection{T4: Adjacency (count connections)}
All participants could correctly identify the number of neighbours of each node with all representations. However, participants were slightly slower with the matrix representation than  with the text representation (see \autoref{tab:time-t4}). \added{Linear mixed modeling revealed a statistically significant effect of representations on time ($\chi^2(3)=7.8, p=.0499$). Post-hoc tests revealed that:} 
\begin{itemize}
    \item Text was faster than Matrix ($p=0.0445$).
\end{itemize}

Self-reported difficulty was similar for all representations (see \autoref{tab:difficulty-t4}). \added{Friedman test revealed no statistically significant effect of representations on difficulty for T4.}

Strategies were similar to those used in task T3. In the text representation participants read along the row to count the number of friends. For the matrix representation they either traced down the column or across the row counting the number of circles.
For the node-link representation participants followed the lines emanating from a node, counting the number of connected nodes.

There was no obvious performance difference between participants with experience of using network data and participants without such experience.

\vspace{-0.5em}
\subsection{T5: Common connection}
Most participants could correctly identify the number of neighbours of each node with all representations. However,
P6 answered this question incorrectly with the matrix. Participants were also significantly  slower with the matrix representation than the text and organic representations (see \autoref{tab:time-t5}). \added{Linear mixed modeling revealed a statistically significant effect of representations on time ($\chi^2(3)=10.1, p=.0181$). Post-hoc tests revealed that:}
\begin{itemize}
    \item Organic was faster than Matrix ($p=0.0445$).
    \item Text was faster than Matrix ($p=0.0642$).
\end{itemize}

The performance on time was in line with self-reported difficulty which was higher for the matrix representation, though this was not statistically significant (see \autoref{tab:difficulty-t5}). 

In the text representation participants read along one row memorising the friends, then follow the second row identifying the friends in common.
 For the matrix representation they either traced down the columns or across the rows in parallel, identifying where they found corresponding circles.
For the node-link representation participants followed links from the first node to all of its neighbours and checked for each neighbour if it was linked to the other node. One participant was at first confused by an edge crossing (P2) but generally edge crossings were understood without difficulty.

Participants with experience of using network data ($M=31s, SD=2.6$) were faster and performed more consistently with Matrix than the participants without experience ($M=84.2s, SD=60.3$).

\begin{table}
    \centering 
    \includegraphics[width=\columnwidth]{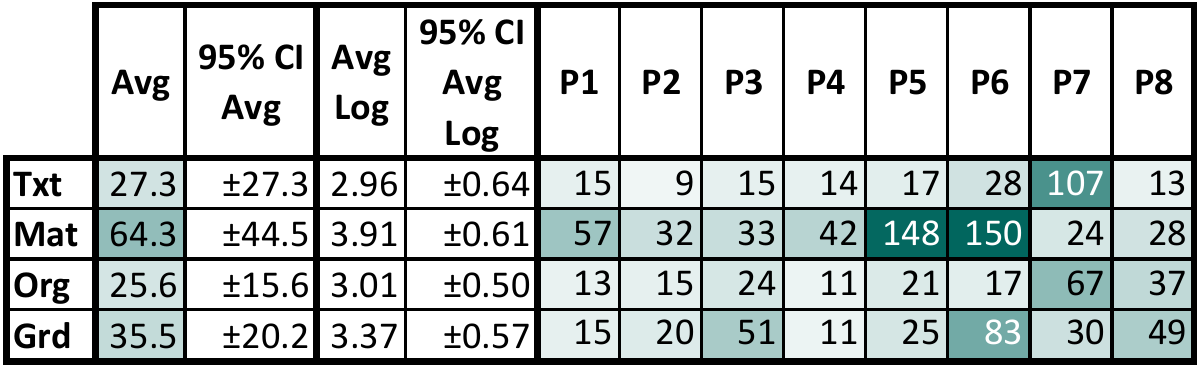}
    \caption{Time for T5: aggregated 95\% confidence interval and time per participant.}
    \label{tab:time-t5}
    \vspace{0.36em}
\end{table}
\begin{table}
    \centering 
    \includegraphics[width=0.9\columnwidth]{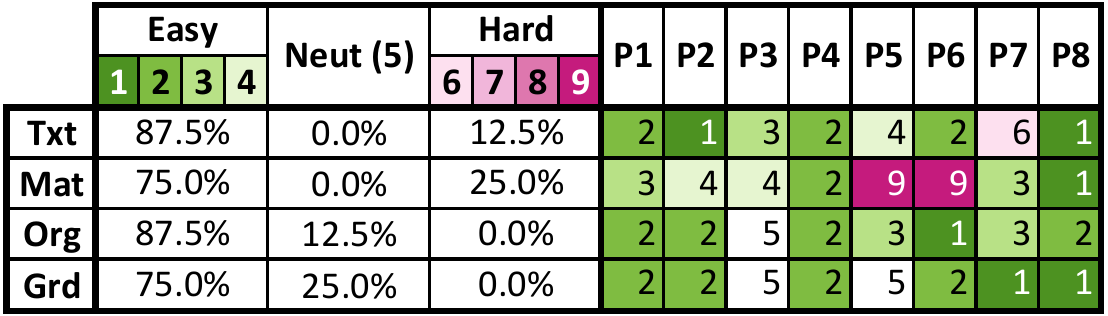}
    \caption{Difficulty for T5: aggregated percentage of difficulty ratings and difficulty ratings per participant.}
    \label{tab:difficulty-t5}
\end{table}

\begin{table}[b!]
    \centering 
    \vspace{-0.5em}
    \includegraphics[width=\columnwidth]{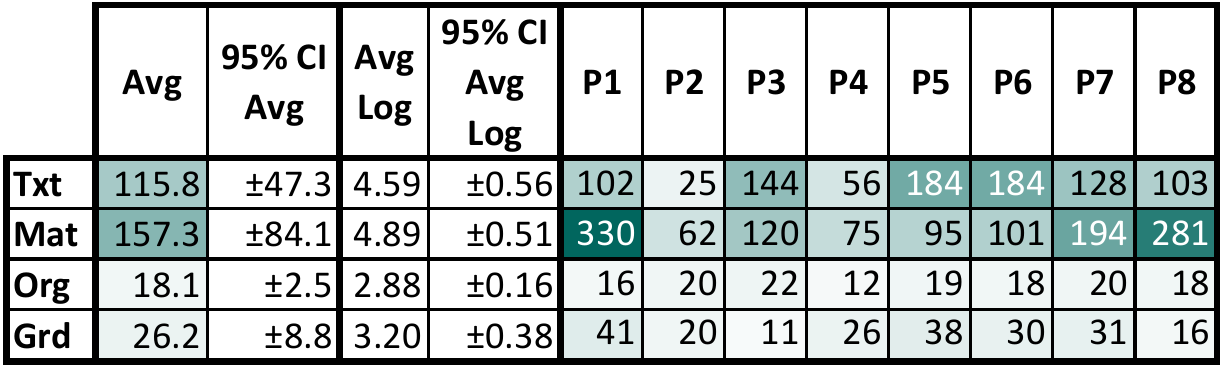}
    \caption{Time for T6: aggregated 95\% confidence interval and time per participant.}
    \label{tab:time-t6}
    \vspace{0.3em}
\end{table}
\begin{table}[b!]
    \centering 
    \includegraphics[width=0.9\columnwidth]{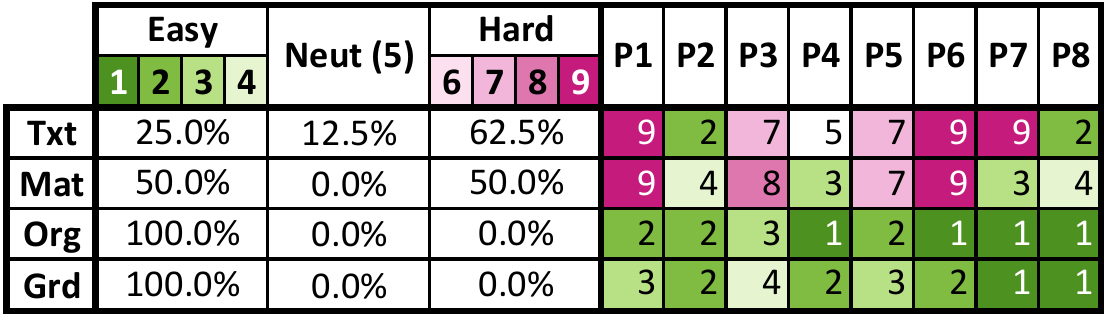}
    \caption{Difficulty for T6: aggregated percentage of difficulty ratings and difficulty ratings per participant.}
    \label{tab:difficulty-t6}
\end{table}

\vspace{-0.5em}
\subsection{T6: Connectivity-path}
This was one of the hardest tasks. P1 and P6 were not able to complete this task with the matrix and P7 was not able to complete this task with the text representation. We see that the node-link representations were significantly faster than text or matrix (see \autoref{tab:time-t6}). \added{Linear mixed modeling revealed a statistically significant effect of representations on time ($\chi^2(3)=46.8, p<.0001$). Post-hoc tests revealed that:} 
\begin{itemize}
    \item Grid was faster than Text ($p<0.0001$) and Matrix($p<0.0001$), and
    \item Organic was faster than Text ($p<0.0001$) and Matrix($p<0.0001$).
\end{itemize}

Node-link representations were also ranked as significantly less difficult (see \autoref{tab:difficulty-t6}). \added{Friedman test revealed a statistically significant effect of representations on difficulty ($\chi^2(3)=20.2, p=.0002$). Post-hoc tests revealed that:}
\begin{itemize}
    \item Grid was easier than Text ($p=0.0843$) and Matrix($p=0.0280$), and
    \item Organic was easier than Text ($p=0.0038$) and Matrix($p=0.0081$).
\end{itemize}

\begin{figure}[t!]
    \centering 
    \includegraphics[width=0.5\columnwidth]{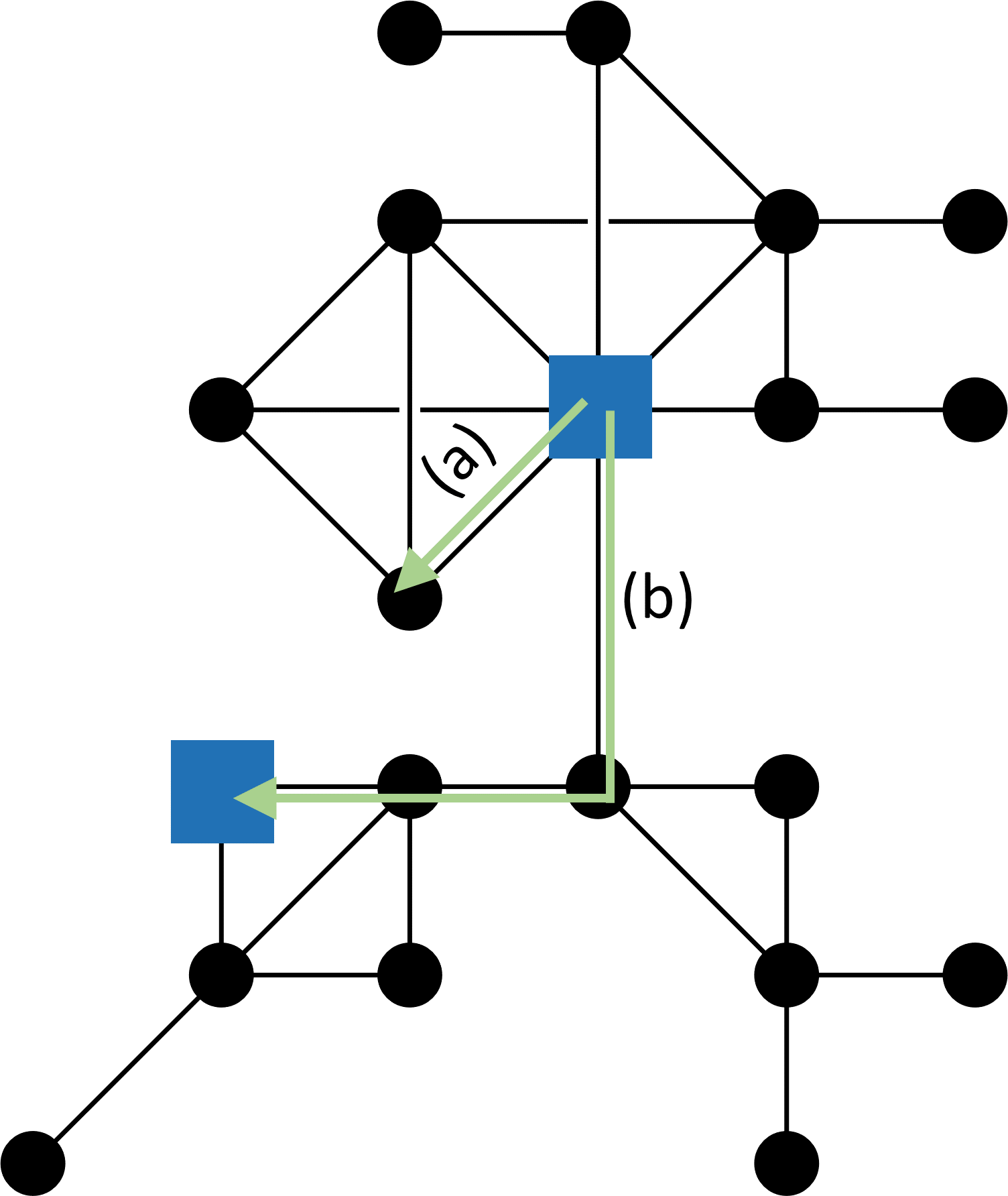}
    \caption{P3 clearly used the geometry of the network to guide her search for a path. She first (a) followed the link directly heading towards the destination node. When she found this was a dead end, she then (b) backtracked and tried the link most closely in  the direction of the destination, following links in this direction until reaching the destination.}
    \label{fig:strategy-t6}
\end{figure}

With the text representation, most participants mentally perform a systematic search through the possible paths, remembering nodes as they encountered them [2,3,4,8]. However some found it too difficult [P7]. They employed a similar strategy with the matrix representation but used the spatial layout of the matrix to guide the search:
\begin{quote}
  \emph{``Look at E and H's friends, remember these, then semi-systematically looking for links between these''} [P3] \newline
  \emph{``I used both the rows and columns to go to the next person''} [P4]  
\end{quote}  
With the node-link representations all participants started from one of the two nodes and semi-systematically followed links that headed in the direction of the other node. \emph{``I followed the direction to the next node.''}[P5] and see for instance \autoref{fig:strategy-t6}.

There was no obvious performance difference between participants with experience of using network data and participants without such experience.

\vspace{-1em}
\subsection{Overall preference}
After completing the tasks with each representation, participants were asked how strongly they agreed that each representation was understandable (S1) and the underlying social network imaginable (S2) (see \autoref{tab:s1} and \autoref{tab:s2}).

Participants agreed significantly more often with both statements with the node-link representations than the text or matrix representation. \added{Friedman test revealed a statistically significant effect of representations on both \emph{understandability} (S1) ($\chi^2(3)=14.8, p=.0020$) and \emph{imaginability} (S2) ($\chi^2(3)=15.3, p=.0015$). Post-hoc tests revealed that:}
\begin{itemize}
    \item Participant agreed more with Grid than with Text (S1: $p=0.0946$; S2: $p=0.0304$), 
    \item Participant agreed more with Grid than with Matrix (S1: $p=0.0028$; S2: $p=0.0732$), 
    \item Participant agreed more with Organic than with Text in S2 ($p=0.0109$), and 
    \item Participant agreed more with Organic than with Matrix (S1: $p=0.0277$; S2: $p=0.0304$). 
\end{itemize}

When asked about what they visualised, one said about the organic representation: \begin{quote}
    \emph{``Liked the representation, imagined the nodes as people. This wasn't true for the braille list: There they were just letters/numbers.''} [P2]
\end{quote} P3 also said that she remembered the text linguistically while for the other representations she remembered them graphically.

They were also asked to rank the representations in terms of \emph{intuitiveness} (R1) and \emph{usability} (R2) (see \autoref{tab:ranking-intuitive} and \autoref{tab:ranking-usability}).

\begin{table}[t!]
    \centering 
    \includegraphics[width=0.98\columnwidth]{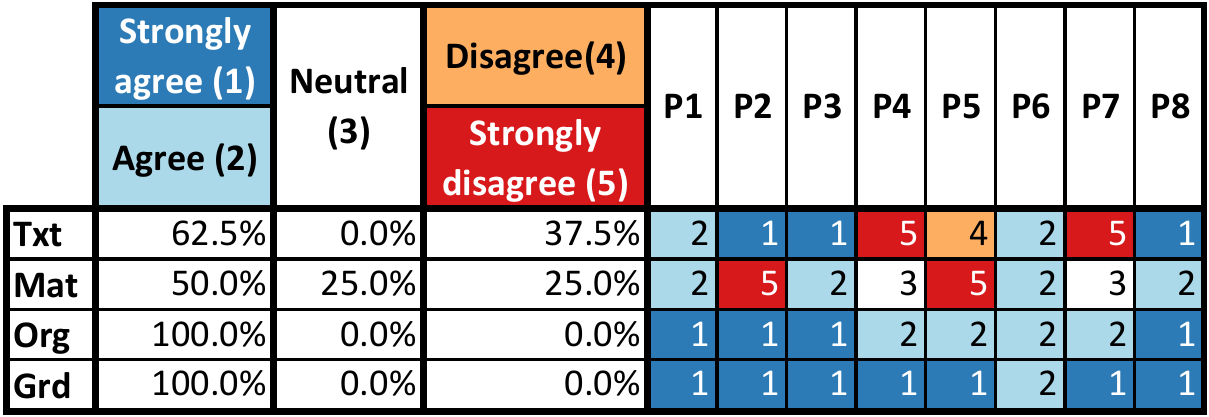}
    \caption{Rating for S1: understandability (\textit{The representation was easy to understand and I barely needed instruction in how to use it.}): aggregated percentage of ratings and ratings per participant.}
    \label{tab:s1}
\end{table}
\begin{table}[t!]
    \centering 
    \includegraphics[width=0.98\columnwidth]{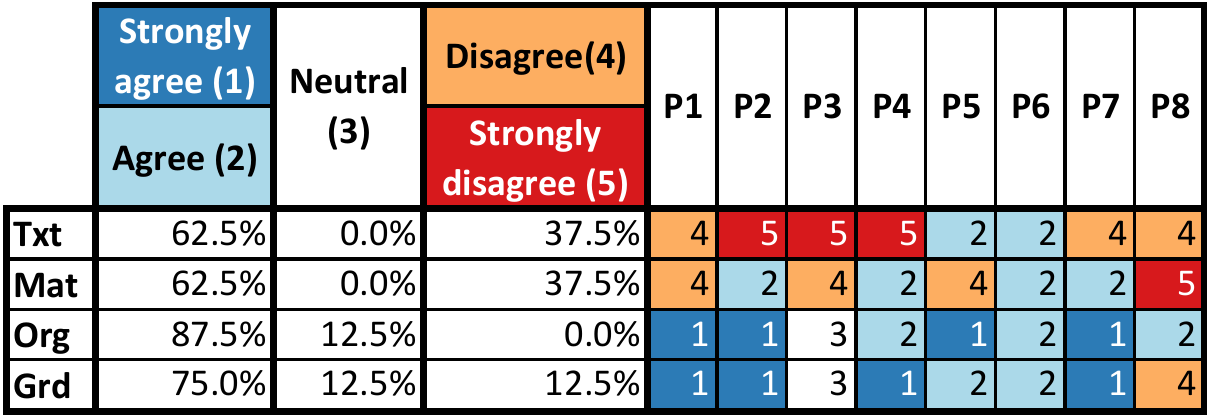}
    \caption{Rating for S2: imaginability (\textit{ I imagined the social network in my head.}): aggregated percentage of ratings and ratings per participant.}
    \label{tab:s2}
\end{table}

\begin{table}[t!]
    \centering 
    \includegraphics[width=0.98\columnwidth]{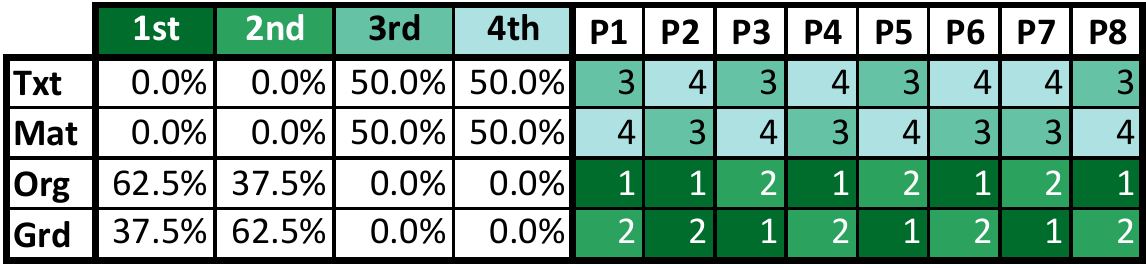}
    \caption{Ranking for R1: intuitive (\textit{The most natural or intuitive way of showing a social network}): aggregated percentage of rankings and rankings per participant.}
    \label{tab:ranking-intuitive}
\end{table}
\begin{table}[t!]
    \centering 
    \includegraphics[width=0.98\columnwidth]{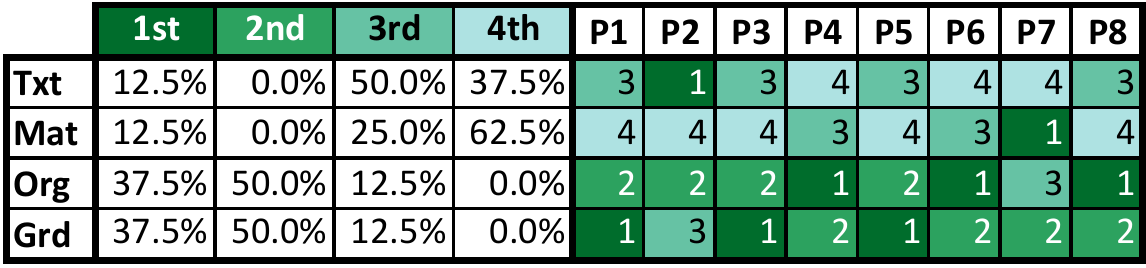}
    \caption{Rating for R2: usability (\textit{In general, the easiest to use}): aggregated percentage of ratings and ratings per participant.}
    \label{tab:ranking-usability}
\end{table}

In line with the above, participants ranked node-link representations higher than the text or matrix representation for both questions. \added{Friedman test revealed a statistically significant effect of representations on both intuitive ($\chi^2(3)=19.4, p=.0002$) and usability ($\chi^2(3)=11.0, p=.0120$). Post-hoc tests revealed that:}
\begin{itemize}
    \item Participant preferred Grid than Text ($p=0.0191$) and Matrix ($p=0.0191$) in terms of intuitive (R1), 
    \item Participant preferred Organic than Text ($p=0.0055$) and Matrix ($p=0.0055$) in terms of intuitive (R1), and 
    \item Participant preferred Organic than Text ($p=0.0572$) and Matrix ($p=0.0572$) in terms of usability (R2). 
\end{itemize}

Comments about the node-link diagrams emphasized that they were easy to understand: 
\begin{quote}
  \emph{`` Love these  [organic] graphics, they make it so easy. This is how I would show someone what a social network is.''} [P1].  
\end{quote}
Some: 
\begin{quote}\emph{``Liked the regularity of the grid - made it predictable. Grid and right angles helped to make [more memorable] mental image}''[P3] 
\end{quote}
while others preferred the organic layout because they felt that the grid layout would not scale to larger networks.

\section{Discussion and Conclusion}

We have described the results of a controlled study with 8 BLV people comparing four different tactile representations of networks. 

All participants found the two node-link diagram representations to be  a more natural and intuitive representation of a social network than an adjacency matrix or textual representation. Participants found  little difference between the organic and grid-layout styles for  node-link diagrams though some participants liked the predictability of the grid layout. Node link diagrams were generally ranked as more usable and resulted in significantly better performance for the two network connectivity tasks--identifying the number of clusters and finding a path between two nodes--than the other two representations. This is in line with results for the equivalent visual representations~\cite{ghoniem2004comparison,keller2006matrices,okoe2018node}. What was surprising is that they also outperformed the matrix and text representation when determining if two nodes are connected.
This difference was also reflected in participant rankings of task difficulty.

Participants performed unexpectedly badly using the matrix representation on the adjacency tasks-finding if two nodes were connected, counting number of neighbours and identifying common connections. In some tasks it performed worse than the text representation. It was clear that participants found it by far the hardest representation to understand. One source of particular difficulty was understanding that the matrix was symmetric and that the diagonal  was the line of symmetry. When viewed visually this symmetry is immediately apparent but not when explored serially by touch. Despite being instructed in this  some participants ignored symmetry leading to over estimates of the number of links and number of clusters, counting those off the diagonal twice.

Studies with sighted participants have not included a textual representation. We found it surprising that the text representation performed surprisingly well, as well as or better than the matrix representation except for the cluster identification task. However, participants comments revealed that it placed a higher demand on working memory than the other representations for identifying clusters, common connections and finding paths. It thus may have difficulty scaling to larger networks. Interestingly,  one participant reported that the textual representation led to a linguistic representation rather than a graphical image. 

Examination of the strategies used by participants to answer the tasks clearly shows that they were taking advantage of the spatial properties of the non-text representations. In all non-text representations the participants used spatial grouping and spatial identification of bridges to determine the number of clusters and in the node-link diagram representations they used the relative position of the two endpoints to guide the search for a path between them. In the matrix representation they also took advantage of row and column alignment to speed the search for mutual friends.

A limitation of the current study is the small sample size. Another is the lack of variability in the graphs that could be tested because of limitations on the length of the study. Further studies should test larger networks and also denser networks as based on studies of visual representations we might expect that the matrix-based representation would outperform the node-link diagram representations. We could also consider alternate less cluttered tactile representations for the matrix or to present a half-matrix. 

Another possible limitation is that we did not include braille labels on the node-link diagrams but did on the text and matrix representation. This was on the advice of our expert transcriber but two participants noted that the node-link diagrams had less information. We do not believe that including braille labels on the nodes would have significantly changed the outcome but this should be tested in a future study.

This research contributes to a more nuanced understanding of the relative advantages of text and graphics to BLV people. 
 We found that touch readers readily extract local information from a tactile graphic, in our case immediate connections between nodes, but for this purpose, there are no benefits over presenting the data in text or list. This accords with prior findings of reading values from tactile bar charts~\cite{goncu2010usability} and line graphs~\cite{aldrich1987tangible}. Touch readers can also use tactile diagrams to support global reasoning, in our case identifying clusters or finding paths. For this purpose, they can be more effective than using text or lists. This accords with previous research on tactile scatter plots~\cite{watanabe2018effectiveness} and extends prior evidence for effectiveness of tactile maps~\cite{ungar1998blind} and tactile statistical graphics~\cite{watanabe2018effectiveness}, to show benefits also hold for more schematic representations.

\section{ACKNOWLEDGMENTS}
Cagatay Goncu is supported by the Australian Research Council (ARC) grant DE180100057. He is also a co-founder of RaisedPixels Pty. Ltd. Leona Holloway is supported by ARC grant LP170100026.

\bibliographystyle{SIGCHI-Reference-Format}
\balance
\bibliography{references}

\end{document}